\documentclass[twocolumn]{aastex631}

\usepackage{newtxtext,newtxmath}

\usepackage[T1]{fontenc}
\usepackage{booktabs}

\DeclareRobustCommand{\VAN}[3]{#2}
\let\VANthebibliography\thebibliography
\def\thebibliography{\DeclareRobustCommand{\VAN}[3]{##3}\VANthebibliography}

\usepackage[absolute,overlay]{textpos}
\usepackage{float}
\usepackage{graphicx}  
\usepackage{subfigure}	
\usepackage{amsmath}	
\usepackage{amssymb}	
\usepackage{color}
\usepackage{xcolor}

\usepackage{enumitem}
\setlist[itemize]{leftmargin=*, labelindent=0pt, itemsep=0ex}

\begin{document}

\title{Identification of Strongly Lensed Gravitational Wave Events Using Squeeze-and-Excitation Multilayer Perceptron Data-efficient Image Transformer}

\author{Dejiang Li}
\affiliation{Department of Physics, and Collaborative Innovation Center for Quantum Effects and Applications, Hunan Normal University, Changsha 410081, China;}
\affiliation{College of Information Science and Engineering, Hunan Normal University, Changsha, Hunan 410081, People's Republic of China;}

\author{Tonghua Liu$^{\star}$}
\affiliation{School of Physics and Optoelectronic Engineering, Yangtze University, Jingzhou, 434023, China;}

\author{Ao Liu}
\affiliation{Department of Physics, and Collaborative Innovation Center for Quantum Effects and Applications, Hunan Normal University, Changsha 410081, China;}
\affiliation{College of Information Science and Engineering, Hunan Normal University, Changsha, Hunan 410081, People's Republic of China;}

\author{Cuihong Wen$^{\ast}$}
\affiliation{Department of Physics, and Collaborative Innovation Center for Quantum Effects and Applications, Hunan Normal University, Changsha 410081, China;}
\affiliation{College of Information Science and Engineering, Hunan Normal University, Changsha, Hunan 410081, People's Republic of China;}

\author{Jieci Wang$^{\ddagger}$}
\affiliation{Department of Physics, and Collaborative Innovation Center for Quantum Effects and Applications, Hunan Normal University, Changsha 410081, China;}

\author{Kai Liao$^{\dagger}$}
\affiliation{School of Physics and Technology, Wuhan University, Wuhan 430072, China;}

\author{Jiaxing Cui}
\affiliation{Department of Physics, and Collaborative Innovation Center for Quantum Effects and Applications, Hunan Normal University, Changsha 410081, China;}
\affiliation{College of Information Science and Engineering, Hunan Normal University, Changsha, Hunan 410081, People's Republic of China;}

\email{$^{\star}$liutongh@yangtzeu.edu.cn;\\$^{\ast}$cuihongwen@hunnu.edu.cn;\\$^{\dagger}$liaokai@whu.edu.cn;\\$^{\ddagger}$jcwang@hunnu.edu.cn}
\begin{abstract}
With the advancement of third-generation gravitational wave detectors, the identification of strongly lensed gravitational wave (GW) events is expected to play an increasingly vital role in cosmology and fundamental physics. However, traditional Bayesian inference methods suffer from combinatorial computational overhead as the number of events grows, making real-time analysis infeasible. To address this, we propose a deep learning model named Squeeze-and-Excitation Multilayer Perceptron Data-efficient Image Transformer (SEMD), based on Vision Transformers, which classifies strongly lensed GW events by modeling morphological similarity between time-frequency spectrogram pairs. By integrating Squeeze-and-Excitation attention mechanisms and multilayer perceptrons , SEMD achieves strong feature extraction and discrimination. Trained and evaluated on simulated datasets using Advanced LIGO and Einstein Telescope noise, the model demonstrates robustness and generalization across different detector sensitivities and physical conditions, highlighting the promise of deep learning for rapid identification of strongly lensed GW signals.
\end{abstract}


\section{Introduction}

Since the first detection of a binary black hole merger event (GW150914) by Laser Interferometer Gravitational-Wave Observatory (LIGO) in 2015~\citep{k1,k38}, ground-based gravitational Wave (GW) detectors have confirmed dozens of binary black hole and binary neutron star mergers. These observations mark the true advent of gravitational wave astronomy and offer new opportunities for testing Einstein's general theory of relativity, including stringent constraints on the propagation speed of GWs and the upper limit of the graviton mass~\citep{k2,k39}. In particular, the GW170817 event—the first binary neutron star merger observed through both gravitational and electromagnetic signals—ushered in the era of multi-messenger astronomy~\citep{k3,k4}. This joint detection provided critical insights into the nuclear physics of neutron star mergers and enabled a measurement of the cosmic expansion rate using the “standard siren” method~\citep{k5}. For example, the binary neutron star merger GW170817 provided the first standard siren measurement of the Hubble constant, yielding a value of $H_0 = 70.0^{+12.0}_{-8.0}\,\mathrm{km\,s^{-1}\,Mpc^{-1}}$, thereby offering an independent and valuable constraint on cosmological models.

When a massive foreground object (such as a galaxy or galaxy cluster) lies between a gravitational wave source and the observer, strong gravitational lensing can occur, producing multiple images of the same GW signal~\citep{k6,2019RPPh...82l6901O,Liao:2022gde}. While this phenomenon has been extensively studied in electromagnetic astronomy, gravitational waves—especially when their wavelength is comparable to the lensing scale—require careful consideration of wave optics effects~\citep{k7}. Under the geometric optics approximation, which assumes the GW wavelength is much smaller than the lens scale, different images from the same source share identical phase evolutions but differ in amplitude. Additionally, the time delays between these images can range from minutes to weeks. Detecting such lensed GW signals not only provides a valuable test of general relativity in the lensing regime but also opens new avenues for astrophysical and cosmological research. For instance, time delays can be used to constrain the Hubble constant~\citep{2017NatCo...8.1148L,k8}, while the statistical properties of lensed events can shed light on the population of lensing galaxies and the distribution of dark matter~\citep{k9}. As detector sensitivity continues to improve, strongly lensed GWs are expected to become a powerful probe of dark matter structures and cosmological parameters.

Traditional approaches for identifying lensed GW events primarily rely on Bayesian model selection, which involves computing the Bayes factor for each candidate event pair to assess whether they originate from the same source, followed by a posterior comparison~\citep{k34,k42}. However, as pointed out by ~\citet{k10}, in the upcoming fourth observing run (O4), hundreds of binary black hole events are expected to be detected, with approximately 1\% potentially being lensed duplicates~\citep{k40}. This would require on the order of $\mathcal{O}(10^2)$ posterior computations and $\mathcal{O}(10^4)$ Bayes factor evaluations. In the era of third-generation detectors, the number of detected events is expected to grow to $\mathcal{O}(10^5$--$10^6)$, with a lensing probability of about 0.3\%, leading to $\mathcal{O}(10^{10}$--$10^{12})$ candidate combinations to be evaluated~\citep{k41}. Such an enormous computational cost renders traditional Bayesian inference impractical for real-time identification of lensed candidates, highlighting the need for more efficient detection strategies.

To address these challenges, recent studies have explored the use of machine learning (ML) techniques for rapid classification of GW events. For example, ~\citet{k10} proposed a method that employs ML models trained on Q-transform spectrograms and sky localization maps, which can be generated within seconds and achieve performance comparable to Bayesian methods, but at a fraction of the computational cost. ~\citet{k11} developed the SLICK deep learning pipeline, which uses parallel networks to analyze Q-transform and sine-Gaussian representations of binary black hole signals. Experiments demonstrate that SLICK achieves a fivefold gain in efficiency at a false positive rate of $10^{-3}$. Overall, deep learning models based on time-frequency representations offer significant advantages: spectrograms and localization maps can be generated almost immediately after detection, whereas full parameter estimation may take hours or days, making these approaches well-suited for large-scale and real-time candidate screening.

In this study, we propose a deep learning model termed Squeeze-and-Excitation Multilayer Perceptron Data-efficient Image Transformer (SEMD), which integrates the Squeeze-and-Excitation (SE) channel attention mechanism~\citep{k12}, multilayer perceptron (MLP)~\citep{k29}, and Data-efficient Image Transformer (DeiT)~\citep{k13}. To accommodate the vertically concatenated structure of time-frequency diagrams induced by lensing, SEMD captures temporal evolution similarities across paired images through DeiT’s cross-image attention modeling. Simultaneously, it employs the SE module to sensitively extract amplitude discrepancies and leverages the nonlinear transformations of MLP to enhance the discrimination of local morphological features. This design is particularly effective for identifying lensed GW events characterized by consistent temporal evolution but differing amplitudes. Compared to conventional single-image classification models, SEMD elevates the detection task from isolated signal identification to relational pattern recognition, aligning more closely with the underlying physics of lensed events.

The remainder of this paper is organized as follows. Section~2 introduces the theoretical background and the procedures for data simulation. Section~3 describes the construction of image samples and the architecture of the proposed SEMD model. Section~4 presents the experimental results and performance evaluations. Section~5 concludes the study and discusses directions for future research.

\section{THEORY}
\label{sec:theory}

Within the standard $\Lambda$CDM cosmological framework, GW signals emitted from distant binary compact objects can be significantly deflected by massive intervening structures such as galaxies or galaxy clusters located along their lines of sight. Under favorable alignments, this results in the formation of multiple images with different arrival times and magnifications—a phenomenon known as strong gravitational lensing~\citep{k14,k43}. The key ingredients in modeling such lensing effects include the lens equation, magnification, and the time delay induced by both geometric and gravitational contributions~\citep{k44}.

In this section, we present the theoretical formalism of gravitational lensing as applied to GWs and describe our simulation pipeline based on the Singular Isothermal Ellipsoid (SIE) lens model, including source and lens population modeling, waveform generation utilizing the IMRPhenomPv2 approximant, and SNR-based filtering~\citep{k45}. The aim is to generate physically interpretable lensed GW signals, incorporating key observables such as image multiplicity, magnification, and time delay~\citep{k15}.

\subsection{Gravitational Lensing of Gravitational Waves}

In the geometric optics limit, the propagation of gravitational wave is modified by the presence of intervening massive structures, such as galaxies or galaxy clusters, along the line of sight~\citep{k46,k47}. When the alignment between the source, lens, and observer is sufficiently close, the gravitational potential of the lens deflects the wavefront, leading to the formation of multiple images. Each image is characterized by a distinct angular position, time of arrival, and amplitude magnification~\citep{k48,k49}. These observable effects are described by the framework of strong gravitational lensing.

The angular positions of the lensed images are governed by the lens equation:
\begin{equation}
\vec{\beta} = \vec{\theta} - \vec{\alpha}(\vec{\theta}),
\end{equation}
where $\vec{\beta}$ denotes the true (unlensed) angular position of the source, $\vec{\theta}$ is the image position, and $\vec{\alpha}(\vec{\theta})$ is the scaled deflection angle determined by the projected surface mass density of the lens. Each physical solution $\vec{\theta}_i$ to Eq.~(1) corresponds to a separate image that may be observed by the detector.

The time delay between each lensed image and the hypothetical unlensed signal consists of both a geometric delay due to the increased path length and a gravitational (Shapiro) delay caused by the time dilation in the gravitational potential of the lens. These contributions are encapsulated in the Fermat potential~\citep{k16}:
\begin{equation}
\phi(\vec{\theta}_i, \vec{\beta}) = \frac{1}{2} |\vec{\theta}_i - \vec{\beta}|^2 - \psi(\vec{\theta}_i),
\end{equation}
where $\psi(\vec{\theta})$ is the two-dimensional lensing potential. The total time delay for the $i$th image is then given by
\begin{equation}
\Delta t_i = \frac{(1 + z_l)}{c} \frac{D_l D_s}{D_{ls}} \, \phi(\vec{\theta}_i, \vec{\beta}),
\end{equation}
where $z_l$ is the lens redshift, and $D_l$, $D_s$, and $D_{ls}$ are the angular diameter distances to the lens, to the source, and between the lens and source, respectively. For certain analytic mass distributions, such as the SIE, the potential $\psi(\vec{\theta})$ can be computed in closed form, enabling efficient evaluation of image properties.

In addition to time delay, each lensed image experiences a magnification determined by the local mapping between the source and image planes. The magnification factor $\mu_i$ is given by the inverse Jacobian determinant of the lens equation:
\begin{equation}
\mu_i = \left| \det \frac{\partial \vec{\beta}}{\partial \vec{\theta}_i} \right|^{-1}.
\end{equation}
This factor quantifies how much the image is stretched (or compressed) relative to the unlensed source. For gravitational waves, whose strain amplitude scales inversely with the luminosity distance, magnification enhances the observed amplitude by a factor of $\mu_i^{1/2}$, thereby increasing the detectability of distant events.

\subsection{Simulation Procedure}

In our simulation, we adopt the SIE model to describe the surface mass distribution of lensing galaxies~\citep{k17,k18}. The SIE profile is a widely used approximation for elliptical galaxies and has been shown to accurately reproduce observed strong lensing systems. It generalizes the Singular Isothermal Sphere (SIS) model by introducing ellipticity in the projected mass distribution.

In this model, the mass distribution is characterized by three primary parameters: the lens redshift $z_l$, the velocity dispersion $\sigma$, and the axis ratio $q$ of the elliptical mass profile. The lens population is sampled consistent with the velocity dispersion function inferred from SDSS data and lensing statistics~\citep{k35,k36}. The SIE model admits analytic expressions for the deflection angle, magnification, and Fermat potential at any image-plane location, which facilitates efficient computation of lensing observables and is well suited for large-scale Monte Carlo simulations. More implementation details are summarized below.

We generate lensed gravitational wave events via a forward Monte Carlo~\citep{k19} simulation pipeline, structured as follows:

\begin{itemize}
  \item \textbf{Source population sampling:} The redshift of the gravitational wave source $z_s$ is drawn from a distribution proportional to $z^2 \exp(-z/0.5)$, truncated at $z_s \leq 5$, which approximates the expected redshift evolution of the merger rate~\citep{k37}. The component masses are sampled independently from a uniform distribution $m_1, m_2 \sim U(5, 50) \,\ M_\odot$.  The dimensionless spin magnitudes are sampled from $U(0,0.99)$, while tilt angles are drawn assuming isotropy i.e., $(\cos\theta \sim U(-1,1))$.  Phase parameters ($\phi_{12}, \phi_{jl}$, coalescence phase), polarization angle ($\psi$), and sky-localization parameters (RA, Dec) are sampled uniformly over their natural domains.  The geocentric coalescence time is fixed at a reference GPS epoch.  The luminosity distance is computed from the sampled redshift using a standard $\Lambda$CDM cosmology.  This parameterization follows the choices made in \citet{k34}.

  \item \textbf{Lensing probability filtering:} We retain only sources with non-negligible probability of multiple imaging by following the optical-depth formalism derived in Appendix~A of Ref.~\citep{k34}. The optical depth can be written as
  \begin{equation}
  \tau(z_s)=\int_0^{z_s} dz_l \; \frac{d\tau}{dz_l},
  \end{equation}
  with
  \begin{equation}
  \frac{d\tau}{dz_l}
  = \int d\sigma \; n(z_l)\,\frac{dp}{d\sigma} \; (1+z_l)^3 \frac{c\,dt}{dz_l}\; \pi D_l(z_l)^2 \, \theta^2(\sigma,z_l,z_s),
  \end{equation}
  where $\tfrac{dp}{d\sigma}$ is the velocity–dispersion function (VDF) of lenses, $n(z_l)$ the comoving number density, and $\theta(\sigma,\cdot)$ the Einstein/angular scale for an SIS/SIE lens. Evaluating this expression under the same assumptions as Appendix~A of \citep{k34} (SDSS VDF, SIS/SIE cross section, chosen cosmology) yields the convenient numerical fit used in our Monte Carlo pipeline:
  \begin{equation}
  \tau(z_s)\approx 4.17\times10^{-6}\left(\frac{D_c(z_s)}{\mathrm{Gpc}}\right)^3,
  \end{equation}
  where $D_c(z_s)$ is the comoving distance to the source. We apply rejection sampling according to this probability to select multiply–imaged candidates.

  \item \textbf{Lens population sampling:} For each selected source, we sample a lens redshift $z_l < z_s$ based on the ratio of comoving volumes. The velocity dispersion $\sigma$ is drawn from a modified Schechter distribution, and the axis ratio $q$ is sampled from a truncated Rayleigh distribution, consistent with galaxy surveys such as SDSS.

  \item \textbf{Image formation and lens equation solving:} We uniformly sample source-plane positions within the strong lensing region of the SIE model. For each lens-source configuration, the lens equation is solved numerically to obtain all physical image positions $\vec{\theta}_i$.

  \item \textbf{Time delay, magnification, and waveform generation:} For each lensed image, the Fermat potential $\phi(\vec{\theta}_i, \vec{\beta})$ is evaluated to compute the corresponding time delay $\Delta t_i$ and magnification $\mu_i$, following Eqs.~(3) and (4) in Section~2.1. These lensing effects are then incorporated into the intrinsic GW signal: the waveform is generated using the IMRPhenomPv2 approximant within the \texttt{bilby} framework, with the effective luminosity distance rescaled as $d_L/\sqrt{\mu_i}$ and the coalescence time shifted by $\Delta t_i$. The resulting lensed waveform is analyzed against a ground-based detector network (H1, L1, V1), and the matched-filter signal-to-noise ratio (SNR) is computed. To ensure observational accessibility, we retain only events with at least two images satisfying $\mathrm{SNR} > 8$.

\end{itemize}

This simulation framework enables the construction of a physically consistent dataset of detectable strongly lensed gravitational wave events, which can be used for training and evaluating classification models.

\section{METHOD}
\subsection{Dataset Description}

To achieve accurate identification of lensed gravitational wave events, we construct two types of datasets for the classification task, denoted as Dataset-L and Dataset-E. The primary difference between the two lies in the background noise: signals in Dataset-L are injected into Gaussian noise generated using the design power spectral density (PSD) of Advanced LIGO~\citep{k20}, while Dataset-E uses the design PSD of the Einstein Telescope (ET)~\citep{k21}. This dual-dataset design aims to evaluate the robustness and generalization ability of the proposed method under the sensitivity conditions of both current and next-generation GW detectors.

\begin{table}[h]
\centering
\caption{Description of the two simulated datasets used in our study.}
\vspace{2mm}
\begin{tabular}{ll}
\toprule
\textbf{Dataset} & \textbf{Description} \\
\midrule
Dataset-L & Simulated GW signals are injected in \textbf{Gaussian} noise, \\
          & generated using Advanced LIGO PSDs. \\
[3pt]
Dataset-E & Simulated GW signals are injected in \textbf{Gaussian} noise, \\
          & generated using Einstein Telescope PSDs. \\
\bottomrule
\end{tabular}
\end{table}

The lensed signals are generated by modeling binary black hole (BBH) mergers under the strong lensing effect, using the SIE model~\citep{k22}. The physical parameters of the source, such as component masses, mass ratio, and inclination, are randomly sampled from predefined astrophysical distributions~\citep{k23}. The unlensed events are standard BBH merger signals not subject to lensing, injected into corresponding Gaussian background noise generated using the PyCBC software package  to ensure dataset completeness\citep{k24}.

All GW signals are generated in the time domain, followed by noise injection and whitening. 
After generating the clean time-domain strain signal $h(t)$, stationary Gaussian noise $n(t)$ consistent with the target detector's design sensitivity is injected. 
The one-sided power spectral density $S_n(f)$ of the detector is used to shape the noise in the frequency domain~\citep{k25}. Specifically, Gaussian random variables are drawn for each Fourier component and scaled by $\sqrt{\tfrac{1}{2} S_n(f)}$ to obtain
\begin{equation}
n(t) = \mathcal{F}^{-1} \left[ \sqrt{\frac{S_n(f)}{2}} \; \xi(f) \right],
\end{equation}
where $\xi(f)$ is a complex Gaussian random variable with zero mean and unit variance, and $\mathcal{F}^{-1}$ denotes the inverse Fourier transform. 
The noisy strain is then given by
\begin{equation}
s(t) = h(t) + n(t),
\end{equation}

To enhance the detectability of the signals, the noisy strain is whitened by dividing its Fourier transform by the square root of the PSD:
\begin{equation}
\tilde{s}_\mathrm{w}(f) = \frac{\tilde{s}(f)}{\sqrt{S_n(f)}},
\end{equation}
followed by an inverse Fourier transform to recover the whitened time-domain waveform:
\begin{equation}
s_\mathrm{w}(t) = \mathcal{F}^{-1} \left[ \tilde{s}_\mathrm{w}(f) \right].
\end{equation}
This whitening process flattens the noise spectrum, ensuring that each frequency contributes equally in matched-filter analyses and improving the visibility of the GW signal in the subsequent Q-transform spectrograms. 
The sampling rate is matched to the PSD used in each dataset, and the whitened time-domain signals are transformed into their corresponding time-frequency representations using the Q-transform, which serve as the input data for the neural network model. 
The details of the image generation and construction procedure are described in Section~3.2.

\subsection{Q-transform and Image Pair Construction}

To convert time-domain gravitational wave signals into formats suitable for deep learning models, we employ the Q-transform~\citep{k26} to generate time-frequency representations. 
The Q-transform is a special case of the continuous wavelet transform (CWT) with a constant quality factor \( Q \), defined as
\begin{equation}
W(a,b) = \int_{-\infty}^{+\infty} h(t) \, \psi^{*}_{a,b}(t) \, dt ,
\end{equation}
where \( h(t) \) is the strain signal, \( \psi^{*}_{a,b}(t) \) is the complex conjugate of the mother wavelet \( \psi(t) \) scaled by factor \( a \) and translated by \( b \). 
For a constant \( Q \), the wavelet is constructed such that the ratio between its central frequency \( f_c \) and bandwidth \( \Delta f \) is fixed, i.e.,
\begin{equation}
Q = \frac{f_c}{\Delta f} .
\end{equation}
In our implementation, we use the complex Morlet wavelet (\texttt{cmor1-3}) with 200 logarithmically spaced scales, corresponding to frequency coverage from 20~Hz to 500~Hz. 
The transform outputs a complex coefficient matrix \( C(f,t) \), whose magnitude 
\begin{equation}
Z(f,t) = | C(f,t) |,
\end{equation}
represents the localized signal energy at frequency \( f \) and time \( t \). 

The resulting two-dimensional spectrogram has time on the horizontal axis, frequency on the vertical axis, and color encoding \( Z(f,t) \) as the instantaneous power spectral density. 
Compared with the short-time Fourier transform, the Q-transform adapts its time-frequency resolution to better capture non-stationary and transient structures in GW signals, making it particularly well-suited for the identification of lensing-induced morphological similarities.

\begin{figure}[htbp]
\centering
\subfigure[Lensed event with Advanced LIGO PSD]{%
    \includegraphics[width=0.48\textwidth]{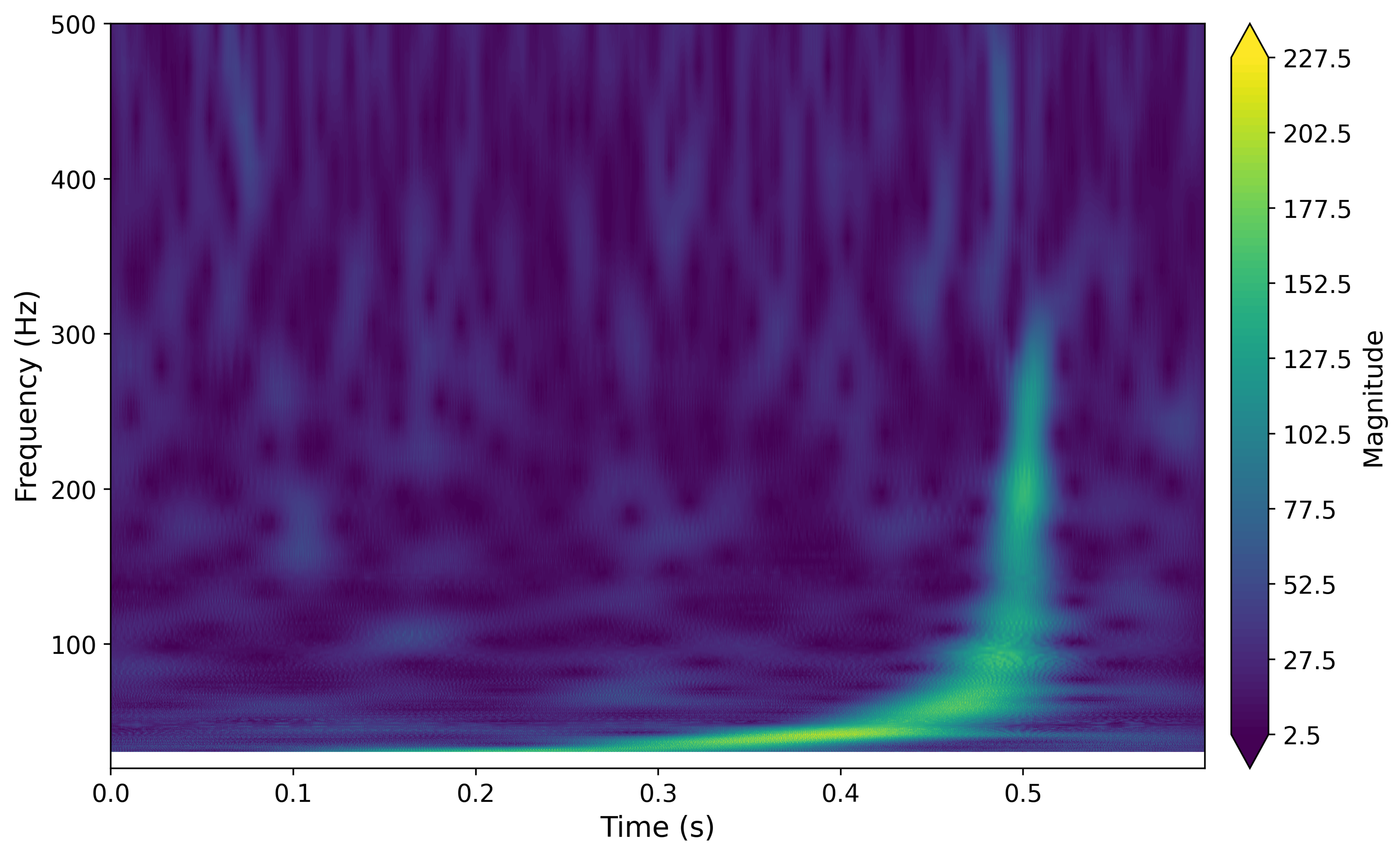}}
\subfigure[Lensed event with Einstein Telescope PSD]{%
    \includegraphics[width=0.48\textwidth]{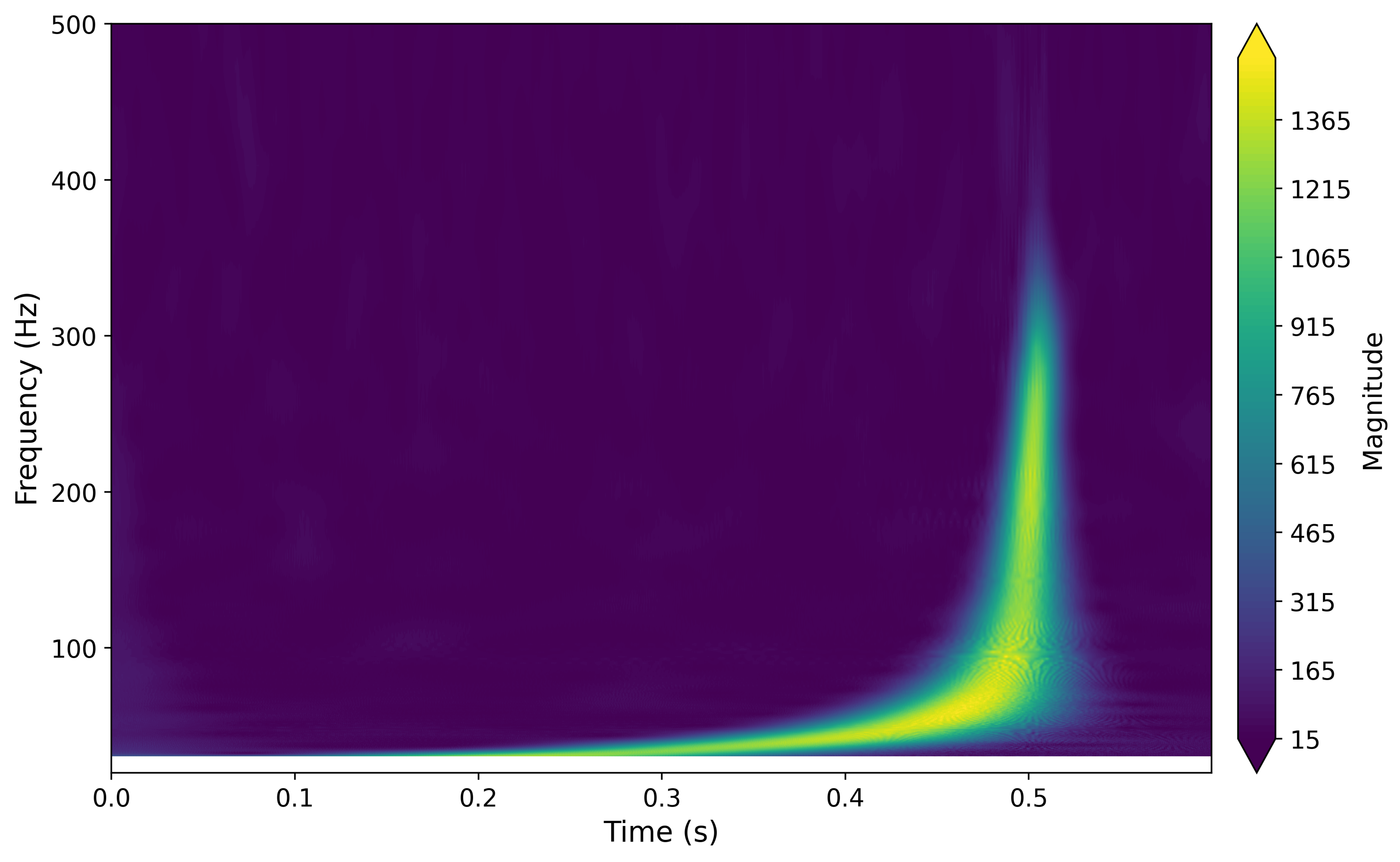}}
\caption{Q-transform spectrograms of lensed gravitational wave signals under different noise conditions. Panel (a) shows a signal injected into Advanced LIGO noise, while panel (b) is under Einstein Telescope sensitivity. Both exhibit similar chirp-like patterns in the time-frequency domain.}
\label{fig:q_transform}
\end{figure}

Figure~\ref{fig:q_transform} shows the Q-transform spectrograms of lensed events injected with Advanced LIGO PSD and Einstein Telescope PSD. Both plots clearly exhibit chirp-like features with increasing frequency, highlighting the energy concentration of compact binary coalescence events. Notably, the spectrogram under Einstein Telescope noise (panel b) appears significantly cleaner and less contaminated by background fluctuations, reflecting the improved sensitivity and reduced noise floor of next-generation gravitational wave detectors. This enhanced clarity allows for more accurate extraction of signal morphology, which is critical for downstream classification \citep{k27}.

In lensed gravitational wave signals, multiple images arise due to strong lensing, typically with slight time delays and magnification differences. These images originate from the same physical event and hence share nearly identical intrinsic parameters\citep{k28}. As a result, their time-frequency representations exhibit similar morphology with different amplitudes. To utilize this morphological similarity in a deep learning framework, we construct image pairs by vertically concatenating two Q-transform spectrograms.

\begin{figure*}[htbp]
\centering
\includegraphics[width=0.9\textwidth]{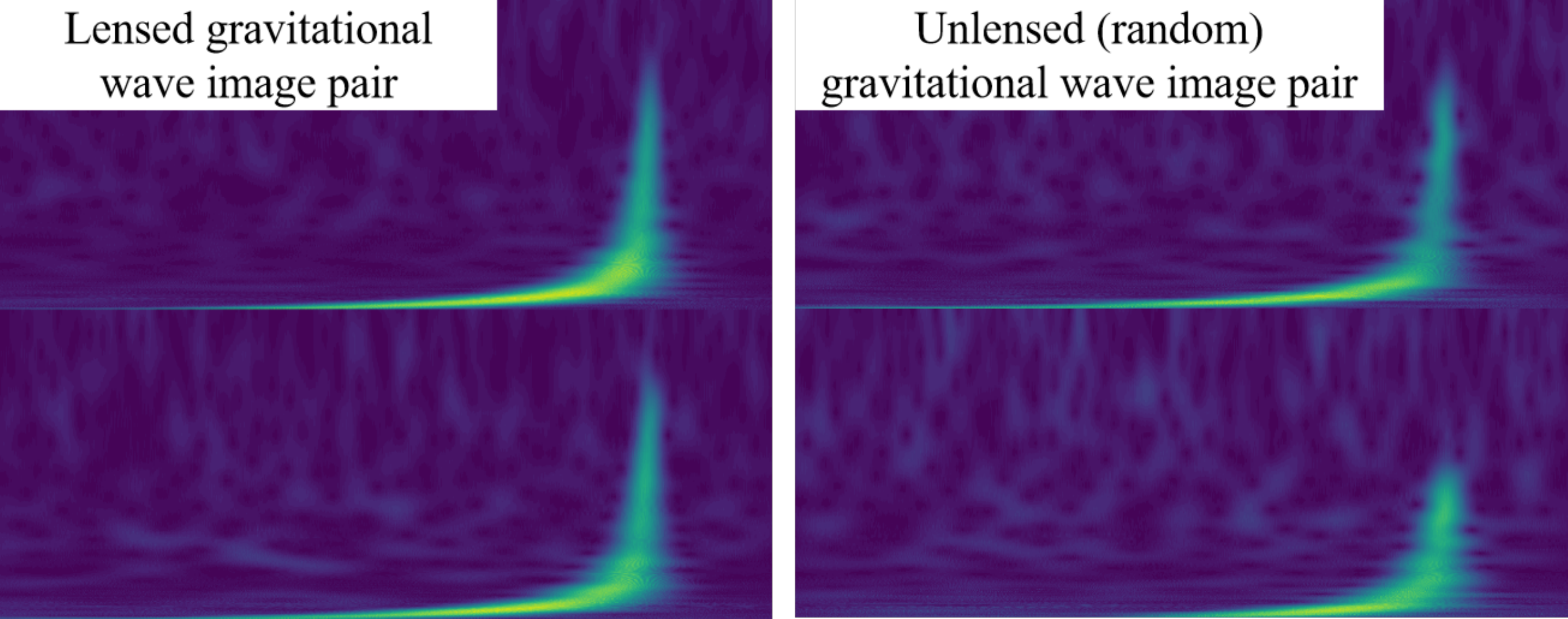}
\caption{Examples of input image pairs used in classification. The left column shows a lensed image pair with similar morphology due to shared intrinsic parameters. The right column shows an unlensed (random) image pair constructed from two independent events, typically exhibiting dissimilar structures. These image pairs are used as direct input to the deep learning classifier.}
\label{fig:pair_comparison}
\end{figure*}

For lensed events, the top and bottom spectrograms in a pair are chosen as the two brightest images from each multiply-imaged event. For unlensed signals, since each event corresponds to a single waveform, two independent events are randomly selected and combined to form a synthetic image pair. This process is illustrated in Figure~\ref{fig:pair_comparison}.

To build the final training dataset, we generate 16,000 lensed image pairs and 16,000 unlensed image pairs for each noise setting (Advanced LIGO and ET). The lensed pairs are constructed as described above, while the unlensed pairs are sampled from different events without lensing. For each noise setting, 8,000 lensed pairs and 8,000 unlensed pairs are used for training, 2,000 lensed pairs and 2,000 unlensed pairs are used for validation, and 6,000 lensed pairs and 6,000 unlensed pairs are used for testing. These paired spectrograms serve as the direct input to the deep learning model.

The lensed image pairs typically show consistent frequency evolution patterns across the top and bottom panels, while unlensed pairs—originating from independent systems with different masses, spins, and orientations—often display distinct structures. This motivates the reformulation of the lensing identification task as a binary image classification problem: determining whether the two parts of a spectrogram pair exhibit morphological similarity. To address this challenge, we propose the SEMD model, which is specifically designed to effectively capture and discriminate these morphological similarities.

All spectrogram images used in this study are subsequently normalized and resized to a consistent resolution before being fed into the neural network model for training and inference. This preprocessing step ensures compatibility with fixed-size input requirements of deep learning architectures and improves training stability.

\begin{figure*}[htbp]
\centering
\includegraphics[width=0.95\textwidth]{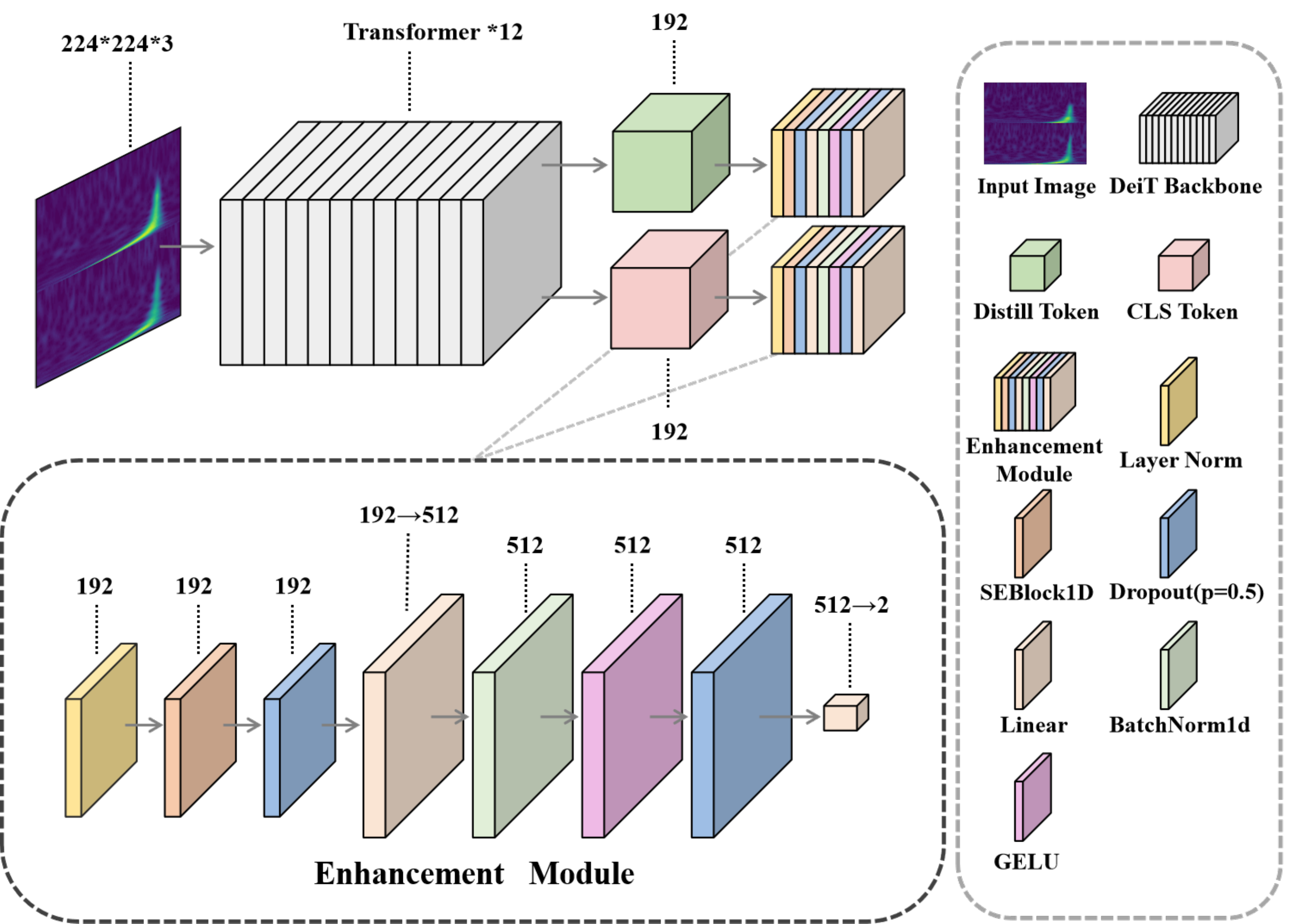}
\caption{Architecture of the proposed SEMD model. It is based on a DeiT-Tiny backbone with dual enhanced heads incorporating Layer Normalization, Squeeze-and-Excitation blocks, and MLP classifiers.}
\label{fig:semd_model}
\end{figure*}

\subsection{Deep Learning Model}

In this study, we propose a novel deep learning model named SEMD, designed for classifying lensed versus unlensed gravitational wave spectrogram pairs. The name SEMD stands for Squeeze-and-Excitation Multilayer Perceptron Data-efficient Image Transformer, reflecting the model’s core components: a Vision Transformer backbone, Squeeze-and-Excitation modules for channel-wise attention, and multilayer perceptrons for classification.

We build SEMD upon the \textit{DeiT-Tiny Distilled Transformer}, a compact yet powerful transformer architecture that employs a two-head design—one classification head and one distillation head~\citep{k13}. This dual-head structure enables auxiliary supervision during training, where the distillation head mimics soft targets from a teacher model, thereby improving data efficiency and generalization~\citep{k30}.

To better suit the lensed image-pair classification task, SEMD introduces architectural enhancements to both heads. Specifically, as illustrated in Figure~\ref{fig:semd_model}, the output embeddings from the transformer backbone are passed through a series of processing layers:

\begin{itemize}
    \item \textbf{LayerNorm}: applied to stabilize the representation from the backbone;
    \item \textbf{Squeeze-and-Excitation Block}: introduces channel-wise attention, enhancing morphological sensitivity to spectrogram structures;
    \item \textbf{Two-layer MLP}: consisting of linear projections, Batch Normalization, GELU activation, and Dropout for effective non-linear transformation and regularization.
\end{itemize}

During training, both the classification head and the distillation head are jointly optimized. Let \( \mathcal{L}_{\text{cls}} \) denote the cross-entropy loss from the classification head, and \( \mathcal{L}_{\text{dist}} \) denote the loss from the distillation head. The overall loss is defined as a weighted combination:
\begin{equation}
\mathcal{L}_{\text{total}} = \alpha \cdot \mathcal{L}_{\text{cls}} + (1 - \alpha) \cdot \mathcal{L}_{\text{dist}},
\end{equation}
where \( \alpha \in [0,1] \) balances the two components. In our experiments, we set \( \alpha = 0.5 \) to ensure equal contribution from both supervision signals.

This combined objective enables the model to benefit from both direct supervision and knowledge distillation. During inference, only the classification head is used for final predictions, ensuring efficiency without compromising accuracy.

\section{RESULT}

To evaluate the performance of the proposed SEMD model under different detector sensitivities, we independently trained two models based on Dataset-L and Dataset-E. Each model was trained on its respective training set, optimized using the validation set, and finally evaluated on the test set, which contains 6,000 lensed and 6,000 unlensed image pairs~\citep{k31}.

\begin{figure*}[htbp]
\centering
\includegraphics[width=0.48\linewidth]{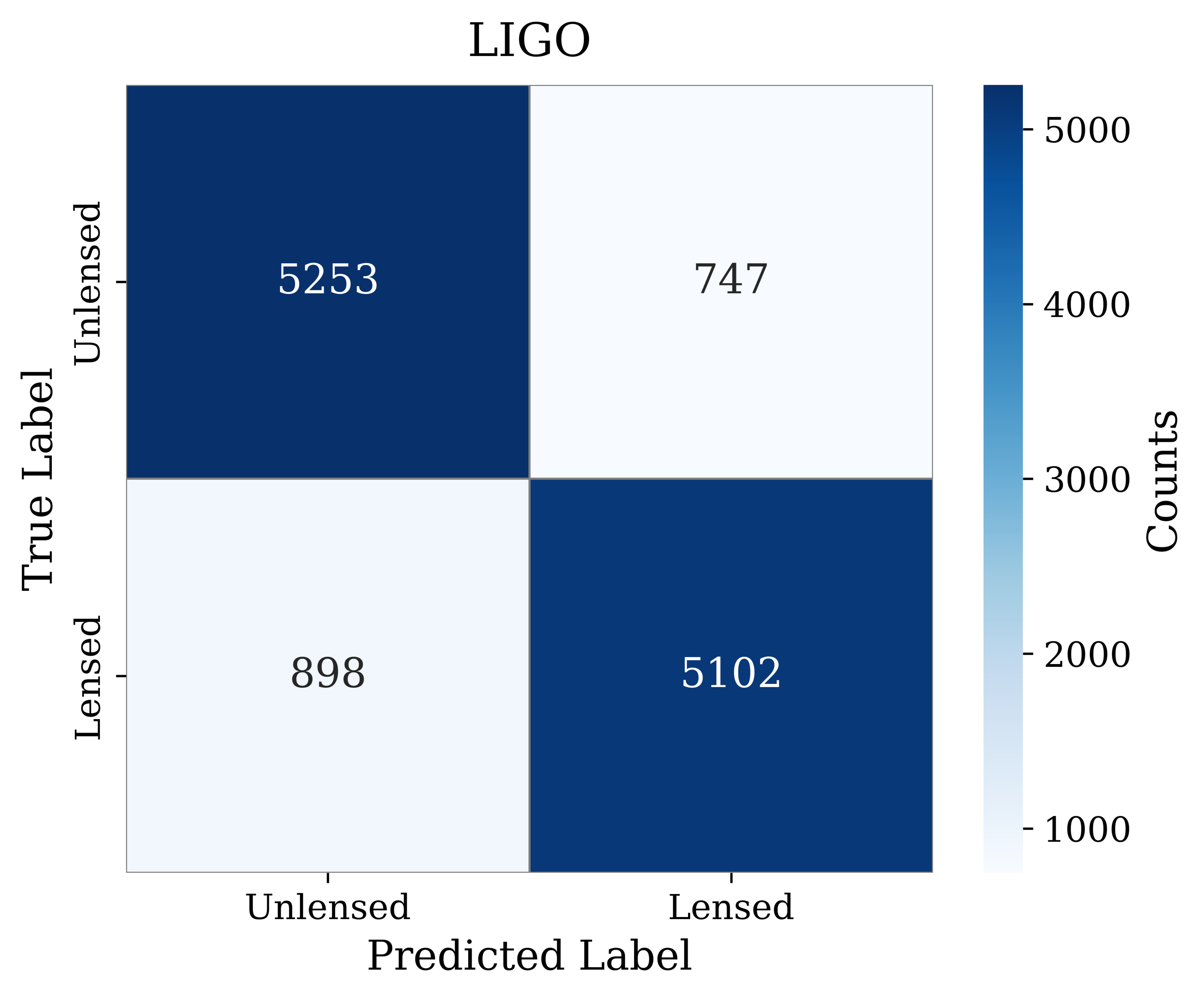}
\includegraphics[width=0.48\linewidth]{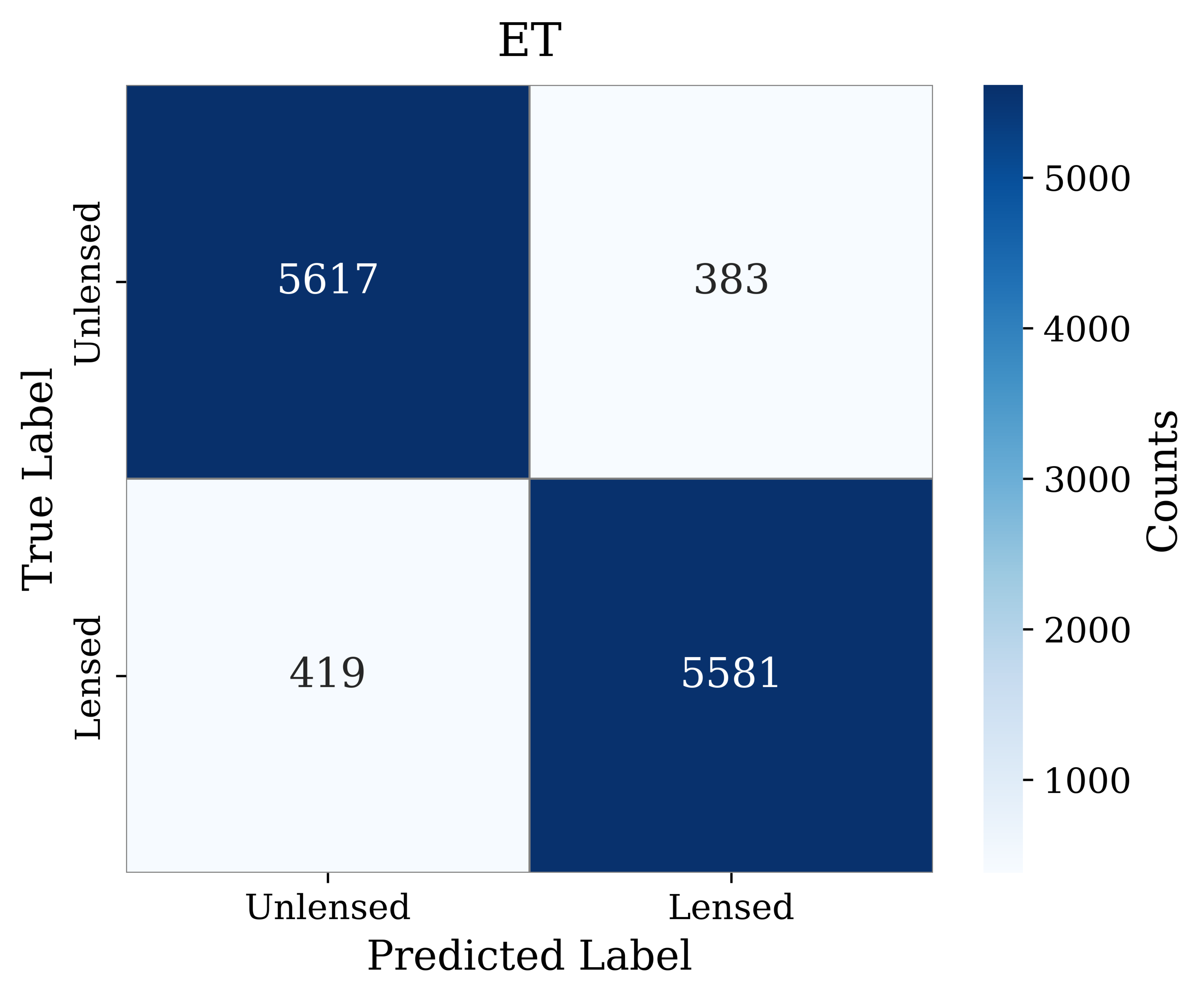}
\caption{ Confusion matrices for SEMD model on different test sets. The model achieves superior classification performance under ET noise conditions, demonstrating the advantage of higher-quality data for lensing identification.}
\label{fig:confusion_matrix}
\end{figure*}

Figure~\ref{fig:confusion_matrix} presents the confusion matrices of the two models on their corresponding test sets. The confusion matrix visualizes the alignment between predicted and true labels, where diagonal elements indicate correctly classified samples, and off-diagonal elements represent misclassifications~\citep{k32}. This allows for a direct assessment of classification accuracy and error distribution. Subfigure (a) shows results for Advanced LIGO noise, while subfigure (b) shows results for Einstein Telescope noise. We find that the model performs better on the Einstein Telescope dataset (Figure~\ref{fig:confusion_matrix}(b), with a lower false positive rate and higher accuracy. This indicates that under conditions of lower noise and higher signal quality, the model can more easily capture the morphological similarities between image pairs.

To further analyze model behavior under varying physical parameters, we partition the test sets based on three attributes:SNR, total binary mass, and mass ratio. We then evaluate the model's performance on each subset using the Receiver Operating Characteristic (ROC) curve, which plots the true positive rate (TPR) against the false positive rate (FPR) under varying classification thresholds~\citep{k33}. The area under the ROC curve (AUC) serves as a standard metric for overall model performance, with larger AUC values indicating better discriminative ability.

To ensure a balanced distribution across subsets, we define the following thresholds for classification:

\begin{itemize}
    \item \textbf{Dataset-L (LIGO PSD):} \\
    $\bullet$ \textit{SNR:} Samples with SNR $>$ 40 are considered high-SNR, others are low-SNR; \\
    $\bullet$ \textit{Total Mass:} The threshold is 60 $M_\odot$; \\
    $\bullet$ \textit{Mass Ratio:} Defined as $q = m_1 / m_2$ with $m_1 < m_2$. Samples with $q > 0.6$ are high mass-ratio, others are low.

    \item \textbf{Dataset-E (ET PSD):} \\
    $\bullet$ \textit{SNR:} A higher threshold of 200 is used due to the generally stronger signals resulting from ET’s improved sensitivity; \\
    $\bullet$ \textit{Total Mass and Mass Ratio:} Same thresholds as in Dataset-L (60 and 0.6, respectively).
\end{itemize}

\begin{figure*}[htbp]
\centering
\subfigure[ROC curve on full test set]{%
    \includegraphics[width=0.49\textwidth]{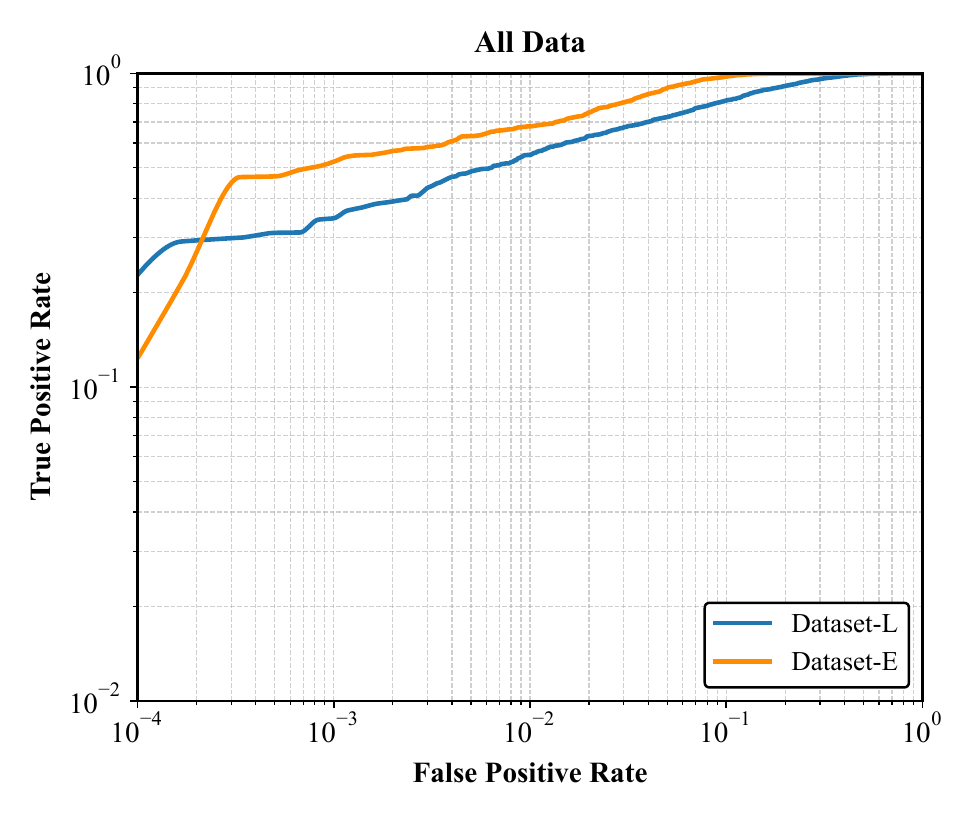}
}
\hfill
\subfigure[ROC curves by SNR]{%
    \includegraphics[width=0.49\textwidth]{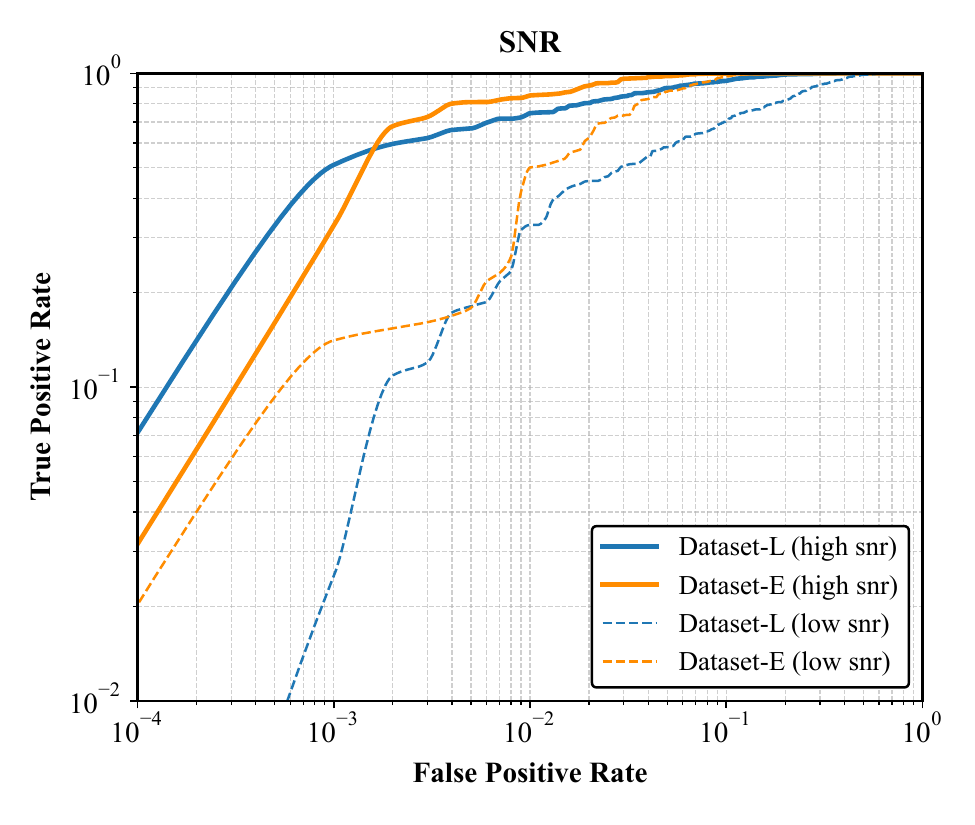}
}
\\
\subfigure[ROC curves by total mass]{%
    \includegraphics[width=0.49\textwidth]{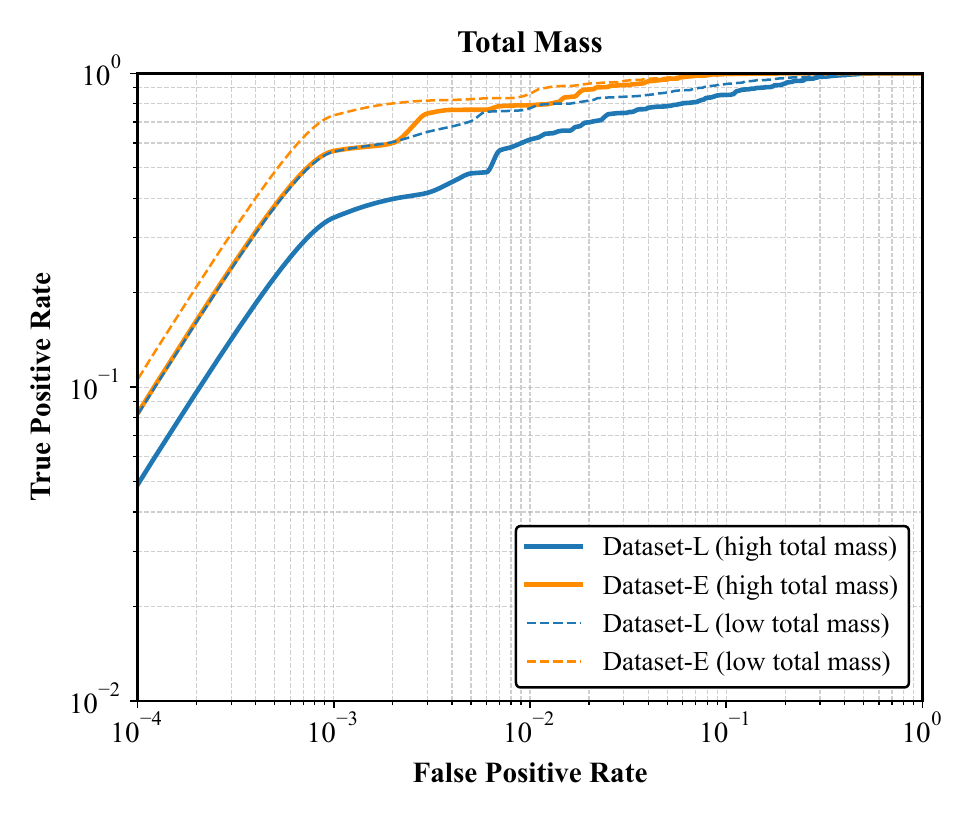}
}
\hfill
\subfigure[ROC curves by mass ratio]{%
    \includegraphics[width=0.49\textwidth]{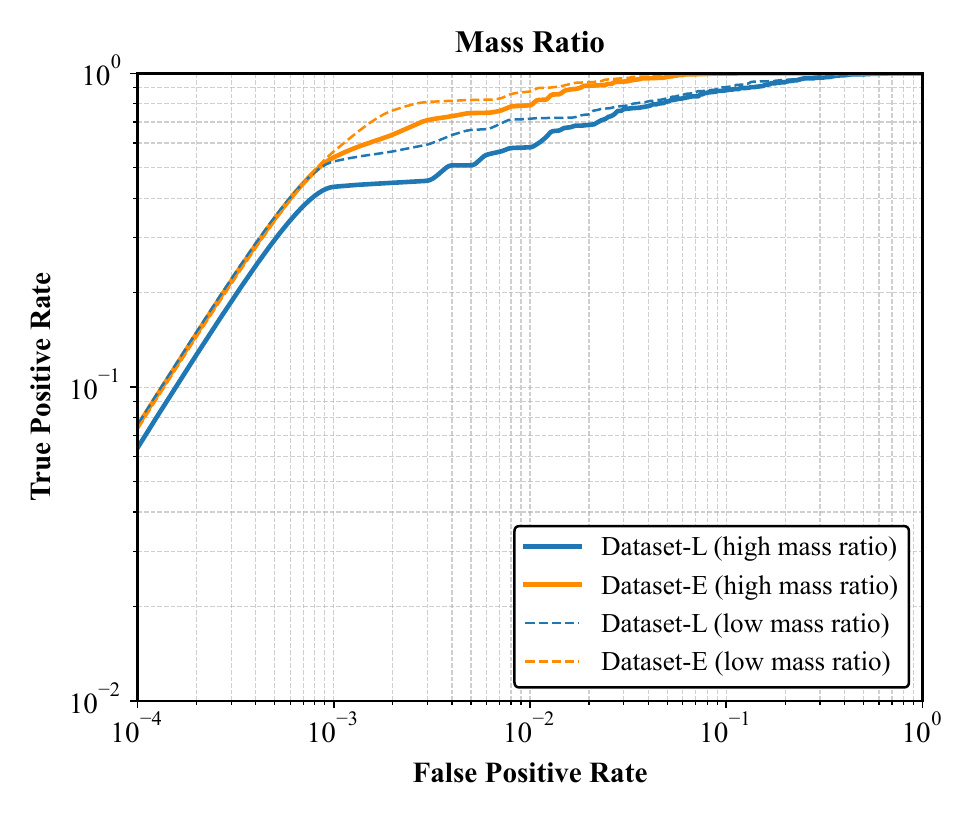}
}
\caption{ROC curves of the SEMD model under different test conditions. Subfigures correspond to: (a) overall test set, (b) signal-to-noise ratio, (c) total binary mass, and (d) mass ratio. The model demonstrates stronger classification capability under ET noise, and better performance on high-SNR, low-mass, and low mass-ratio subgroups.}
\label{fig:roc_analysis}
\end{figure*}

Figure~\ref{fig:roc_analysis} displays the ROC curves under each scenario. Subfigure (a) presents the ROC curve on the complete test set. Subfigures (b), (c), and (d) show results for high/low SNR, high/low total mass, and high/low mass ratio groups, respectively.

In each plot, blue lines represent Dataset-L, and orange lines represent Dataset-E. Thick solid lines indicate high-SNR, high-mass, or high mass-ratio groups, while thin dashed lines indicate their corresponding low-value groups. This consistent visual encoding allows direct comparisons between different physical conditions and detector sensitivities.

Several key observations emerge:

\begin{itemize}
    \item The SEMD model achieves higher AUC scores on Dataset-E compared to Dataset-L, reaffirming the advantage of low-noise, high-fidelity data from next-generation detectors like ET.
    \item For both datasets, high-SNR samples yield better performance, consistent with the improved clarity and signal structure aiding morphological analysis.
    \item Interestingly, the model performs better on low total mass samples, likely because lower mass binaries generate longer-duration waveforms with more distinct time-frequency evolution.
    \item Samples with low mass ratio (more asymmetric systems) also achieve higher AUC, suggesting stronger distinguishable features in their spectrogram morphology.
\end{itemize}

Beyond classification accuracy, the practicality of a model for large-scale gravitational wave searches also depends critically on its inference efficiency and computational resource requirements. 
Once trained, the SEMD model is capable of processing approximately ten thousand paired spectrograms in about two minutes on a single general-purpose GPU. 
In contrast, traditional Bayesian inference methods, even after obtaining parameter estimation posteriors, typically require several additional minutes to evaluate the Bayes factor for a single candidate pair, while generating the posteriors themselves may take from several hours to multiple days depending on the event complexity and computational resources~\citep{k50}. 
This stark difference highlights the substantial efficiency advantage of the proposed approach, enabling timely screening of vast numbers of candidate pairs and supporting near real-time follow-up analyses. 
Moreover, the model’s lightweight architecture ensures low GPU memory consumption and minimal CPU usage, making it suitable for integration into existing GW detection pipelines without the need for dedicated high-performance computing clusters. 
Such characteristics are particularly advantageous for upcoming third-generation detectors, which are expected to deliver orders-of-magnitude increases in event rates.

\begin{table}[htbp]
\centering
\caption{Inference efficiency and resource usage of the trained SEMD model (batch size = 1024). 
Measurements are averaged over multiple runs.}
\vspace{2mm}
\setlength{\tabcolsep}{4pt} 
\renewcommand{\arraystretch}{1.1} 
\begin{tabular}{lcc}
\toprule
\textbf{Metric} & \textbf{Dataset-L} & \textbf{Dataset-E} \\
\midrule
Time to process $10^4$ pairs & $\sim$ 125 s & $\sim$ 120 s \\
Estimated throughput (pairs/s) & $\sim$ 80 & $\sim$ 83 \\
GPU memory usage & $<$ 4 GB & $<$ 4 GB \\
CPU usage (single thread) & $<$ 15\% & $<$ 15\% \\
\bottomrule
\end{tabular}
\label{tab:inference_efficiency}
\end{table}
Given the expected detection rates of $10^5$--$10^6$ binary merger events per year by third-generation detectors, such as the Einstein Telescope and Cosmic Explorer, the ability of the SEMD model to rapidly and efficiently process large volumes of candidate pairs becomes particularly valuable~\citep{k51,k52}. 
Its high-throughput inference allows for near real-time identification of strongly lensed events from massive data streams, ensuring that subsequent Bayesian parameter estimation and multi-messenger follow-ups can be promptly targeted at the most promising candidates. 
This capability not only alleviates the computational bottlenecks inherent in traditional Bayesian pipelines but also significantly enhances the timeliness and scientific return of future gravitational wave observing runs.

\section{SUMMARY}
As the number of gravitational wave detections continues to increase rapidly, traditional Bayesian inference-based methods for identifying strongly lensed events are facing significant computational challenges. These approaches typically require posterior reconstruction and Bayes factor evaluation for every event pair, resulting in a computational cost that scales quadratically with the number of events—posing serious limitations for real-time analysis in the era of third-generation detectors. Moreover, such methods are often sensitive to prior assumptions and struggle to adapt to diverse noise environments and complex physical scenarios. This highlights the urgent need for an efficient, automated, and physically interpretable alternative.

To address these challenges, we propose a novel deep learning framework named SEMD for the identification of gravitational wave events affected by strong lensing. By framing the task as a morphological similarity classification problem between pairs of time-frequency images, SEMD aims to replace the high computational burden of traditional Bayesian inference with a scalable and efficient candidate filtering scheme.

We construct two physically realistic datasets using power spectral densities from different detectors: Dataset-L and Dataset-E . Each dataset is generated through a complete physical modeling pipeline, including source parameter sampling, strong lensing simulation using the SIE model, image construction, and SNR-based selection. Lensed image pairs are formed by the two brightest images originating from the same source event, while unlensed image pairs are randomly assembled from independent events.

In terms of architecture, SEMD integrates the global modeling capabilities of the lightweight DeiT-Tiny Transformer, incorporates a Squeeze-and-Excitation block to enhance channel-wise attention to morphological differences, and employs a multi-layer perceptron classification head for nonlinear transformation and decision-making. During training, we apply a weighted combination of classification and distillation head losses, while inference uses only the output of the classification head to ensure both accuracy and efficiency.

We trained and evaluated the SEMD model separately on both datasets. Test results demonstrate that SEMD achieves robust classification performance under both detector conditions, with higher accuracy and fewer misclassifications observed on the Einstein Telescope dataset. This suggests that high-fidelity signals enable the model to more effectively extract morphological features. To further assess the model's performance under different physical conditions, we partitioned the test samples by SNR, total mass, and mass ratio, and conducted quantitative analysis using ROC curves. We find that the model consistently maintains strong discriminative performance across all subsets, demonstrating excellent adaptability and generalization capability.

In future work, we plan to incorporate multi-detector joint analysis, validate the model under realistic non-Gaussian noise conditions, and explore integration with multimodal information such as sky localization and distance estimates. These efforts aim to establish a more accurate and real-time-capable framework for identifying strongly lensed gravitational wave events.

\section*{Acknowledgements}
This work was supported by National Key R$\&$D Program of China (No. 2024YFC2207400), and was supported by the National Natural Science Foundation of China under Grants No. 12203009, 12374408  and 12475051;  the science and technology innovation Program of Hunan Province under grant No. 2024RC1050;.

\bibliography{sample631.bib}{}
\bibliographystyle{aasjournal}

\end{document}